\documentclass[aps,prb,10pt,twocolumn,raggedbottom,nobalancelastpage]{revtex4-1}
\usepackage{graphicx}
\usepackage{amsmath}
\usepackage{braket}

\newcommand{\dvec}[1]{\ensuremath{\boldsymbol{#1}}}
\newcommand{\vk}{\dvec{\mathrm{k}}}
\newcommand{\vq}{\dvec{\mathrm{q}}}

\begin{document}
\title{Optical and transport gaps in gated bilayer graphene}
\author{Hongki Min}
\author{D. S. L. Abergel}
\author{E. H. Hwang}
\author{S. Das Sarma}
\affiliation{Condensed Matter Theory Center, Department of Physics,
University of Maryland, College Park, Maryland 20742, USA }

\begin{abstract}
We discuss the effect of disorder on the band gap measured in bilayer
graphene in optical and transport experiments. By calculating the
optical conductivity and density of states using a microscopic model
in the presence of disorder, 
we demonstrate that the gap associated with transport experiments is
smaller than that associated with optical experiments.
Intrinsic bilayer graphene has an optical conductivity in which the
energy of the peaks associated with the interband transition are very
robust against disorder and thus provide an estimate of the band gap.
In contrast, extraction of the band gap from the optical conductivity of 
extrinsic bilayer graphene is almost impossible for significant levels
of disorder due to the ambiguity of the transition peaks.
The density of states contains an upper bound on the gap measured in
transport experiments, and disorder has the effect of reducing this gap
which explains why these experiments have so far been unable to
replicate the large band gaps seen in optical measurements.
\end{abstract}

\maketitle

Study of transport in graphene and related materials has recently
become the focus of intensive research efforts since the experimental
realization of gated samples in 2005 \cite{castroneto2009, 
dassarma-rmp, abergel2010}.
Bilayer graphene (BLG) consists of two parallel sheets of graphene
coupled in the Bernal (AB) stacking arrangement, and the low-energy
electronic quasiparticles behave like massive chiral fermions.
This system shows unique physics, with many different physical effects
competing to dominate the observed properties and this has made BLG
a very attractive material to study in the pursuit of
understanding of fundamental physics of chiral materials.
One of the most appealing features of this system is the dynamically
tunable band gap which may be opened and controlled by a perpendicular
electric field \cite{mccann2006, min2007}.
This is the most prominent reason for the explosion of interest in
BLG because the potential applications of a tunable narrow
gap semiconductor are legion.
However, the topic of charge transport through the gapped bilayer system
is still a controversial issue. 

Various experiments have examined transport
\cite{oostinga2007, zou2010, taychatanapat2010, xiao2010, xia2010,
yan2010}
and the optical properties
\cite{ohta2006, zhang2009, wang2008, kuzmenko2009, mak2009, li2009}
of biased bilayer graphene.
In many cases, the band gap extracted from a transport
measurement was many times smaller than that found optically.
The most obvious explanation for this discrepancy is sample disorder
such as charged impurities and short-range scatterers
\cite{adam2007,dassarma2010}.
Some theoretical work on the density of states (DOS) and optical
conductivity of biased BLG in the presence of disorder 
using coherent potential approximation \cite{nilsson2007},
in the instanton approach to the in-gap fluctuation states
\cite{mkhitaryan2008}, for lattice defects \cite{peres2006}, for midgap
states and Coulomb scatterers \cite{stauber2008}, and for finite-ranged
scatterers \cite{ando2011} has been published, but to this date no
theoretical studies exist which consider the issue of transport and
optical gaps in BLG in the presence of disorder.

In this Rapid Communication, we present a comprehensive theory of the
optical conductivity in gapped BLG and the associated
DOS. From these quantities we can describe the gaps
measured by optical and transport measurements and make direct
comparisons between them for the same system. The optical gap is best
found from intrinsic graphene (where there are no excess charge
carriers) because the relevant features in the optical conductivity are
much more robust against disorder in that case. In contrast, the
measurement of the band gap in extrinsic graphene (\textit{i.e.} BLG
with finite carrier density) relies on spectral features which
are strongly affected by disorder. One of the main achievements of this
work is the development of a quantitative microscopic theory for the
optical conductivity of BLG in the presence of disorder,
and the direct comparison with the DOS at the same level
of approximation. We also demonstrate for the first time why the gaps
extracted via different experimental techniques vary so widely. Finally,
we shall propose that manufacturing very clean samples will lead to much
closer agreement in the optical and transport gaps.

To accomplish this, we compute the DOS and optical
conductivity including disorder within the self-consistent Born
approximation (SCBA). The results will be similar for various
scattering mechanisms, but we choose Coulomb scatterers with
Thomas-Fermi screening \cite{hu2008, dejuan2010} to illustrate them.
Therefore, the impurity density is a parameter which describes the
effective disorder concentration, and not strictly the density of
charged impurities.

In the SCBA, the self-energy is computed as follows:
\begin{equation}
	\Sigma_{\lambda}(\vk,E) = 
	 \sum_{\lambda'} \int \frac{d^2 \vk'}{(2\pi)^2} \\ 
	\frac{ n_{\rm imp} \left| V_\mathrm{imp}(\vk-\vk') \right|^2 
		F_{\lambda \lambda'}(\vk,\vk')}
	{E-E_{\lambda' \vk'} - \Sigma_{\lambda'}(\vk',E) + i\eta}.
\end{equation}
Here $\lambda$ is the band index, $n_\mathrm{imp}$ is the impurity
density, $\eta$ is an infinitesimally small positive number, 
$F_{\lambda \lambda'}(\vk,\vk')$ is square of the wave function overlap
between $\ket{\lambda,\vk}$ and $\ket{\lambda',\vk'}$ states, which is a
nontrivial function of the wave vectors and band indices. The Fourier
transform of the impurity
potential is denoted by $V_\mathrm{imp}(\vq)$ at wave vector $\vq$. 
This self-energy can be used to compute the retarded electron Green's
function $G_{\lambda}(\vk,E) = \left(E - E_{\lambda \vk} -
\Sigma_{\lambda}(\vk,E) + i\eta\right)^{-1}$ in the presence of
disorder, from which the DOS can be extracted in the usual way:
\begin{equation}
	D(E)= -\frac{g_{\rm s} g_{\rm v}}{\pi} \sum_{\lambda} 
	\int \frac{d^2 \vk}{(2\pi)^2} 
	\mathrm{Im} \left[G_{\lambda}(\vk,E)\right].
\end{equation}
This includes the spin and valley degeneracy factors $g_\mathrm{s} =
g_\mathrm{v} = 2$ since the charged impurity scattering does not mix
valleys or spins.

For the screened Coulomb interaction with the effective impurity
distance $d_{\rm imp}$ from the graphene layer, 
$V_{\rm imp}(\vq)$  is given by
\begin{equation}
	\label{eq:screen}
	V_\mathrm{imp}(\vq) = 
		\frac{2\pi e^2}{\epsilon_0 \left(q+q_{\rm s}\right)}
		e^{-q d_{\rm imp}}, 
\end{equation}
where $\epsilon_0$ is the background dielectric constant and $q_{\rm s}$
is the screening wave vector. 
For a finite carrier density (the extrinsic case), the screening
wave vector can be approximated by $q_{\rm s}\approx q_{\rm TF}$ where
$q_{\rm TF}= \frac{2\pi e^2}{\epsilon_0} D_0(E_{\rm F})$ is the
Thomas-Fermi wave vector and $D_0(E_{\rm F})$ is DOS at the Fermi energy
$E_{\rm F}$ in the absence of disorder.
However, if the Fermi energy lies in the gapped region (\textit{i.e.} for
intrinsic graphene), there is no DOS at the Fermi energy and the
screening wave vector should be defined by the interband contribution to
the polarization. These two methods of defining the screening 
wave vectors provide similar screening lengths \cite{abergelunpub}.

The optical conductivity is computed within the Kubo formalism
\cite{mahan2000} and at zero temperature can be summarized by the
following expression:
\begin{multline}
	\sigma_{xx}(E) = 2 g_s g_v \frac{e^2}{h} \sum_{\lambda,\lambda'} 
	\int \frac{k'dk'}{2\pi} \int_{E_{\rm F}-E}^{E_{\rm F}} \frac{dE'}{E} \\
	\times M_{\lambda\lambda'}^2(k') 
	{\rm Im}G_{\lambda}(k',E'){\rm Im}G_{\lambda'}(k',E'+E).
	\label{eq:conductivity}
\end{multline}
where
\begin{equation}
	M_{\lambda\lambda'}^2(k) = \int_0^{2\pi} \frac{d\phi}{2\pi}
	\left|\bra{\lambda,k,0} \hbar \hat{v}_x \ket{\lambda',k,\phi)}
	\right|^2
\end{equation}
and
$\phi$ is the angle of the wave vector $\vk$.
Throughout this Rapid Communication, we use the four band tight-binding
Hamiltonian \cite{dassarma2010, abergel2010} of
BLG which includes the intralayer nearest-neighbor hopping
elements parametrized by $\gamma_0=3$ eV, the hopping via
interlayer dimer bonds $\gamma_1 = 0.3$ eV, and the onsite
potential energy supplied by the gates given by $\pm u/2$ depending on
which layer contains the lattice site\cite{abergel2010}.  

\begin{figure}[tb]
	\includegraphics[width=0.98\linewidth]{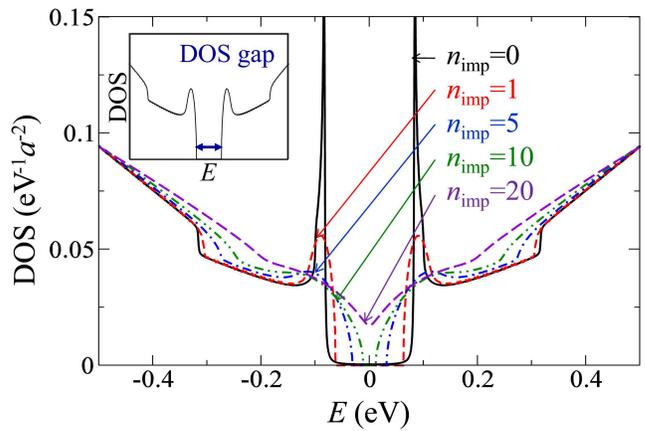}
	\caption{(Color online) Density of states (DOS) of intrinsic biased
	BLG with $u=0.2$ eV on SiO$_2$ substrate.  The 
	case with no charged impurities is shown by the black line.  The
	other lines show the DOS for increasing charged impurity
	concentration, with $n_{\rm imp}$=1, 5, 10, 20$\times
	10^{12}\,\mathrm{cm}^{-2}$.}
	\label{fig:dos}
\end{figure}

We can now describe the transport and optical gaps within the framework
of our theory, defining the gaps as follows.
We characterize the gap in the absence of disorder by the associated
potential energy difference between the layers in the single particle
tight-binding formalism, and label this quantity by $u$.
The ``sombrero'' dispersion of gapped BLG \cite{mccann2006}
means that the minimum band gap occurs slightly away from the K point
such that the bare band gap is
$E_\mathrm{gap}^{(0)}=u\gamma_1/\sqrt{\gamma_1^2+u^2}$.  
The DOS gap is a theoretical quantity defined as the energy difference
between the valence and conduction bands as defined by the onset of the
DOS in the SCBA calculation (see the inset to Fig.
\ref{fig:dos}), while the transport gap is extracted from experiments.
We note that, due to the possible existence of mid-gap impurity states
as well as puddle formation \cite{rossi2011}, the transport gap may be
lower than the DOS gap. 
To the best of our knowledge, no evidence of strong localization has so
far been seen in transport experiments on BLG, 
and this allows us to assume that the DOS gap is the effective upper
bound for the transport gap.

\begin{figure}[tb]
	\includegraphics[width=0.98\linewidth]{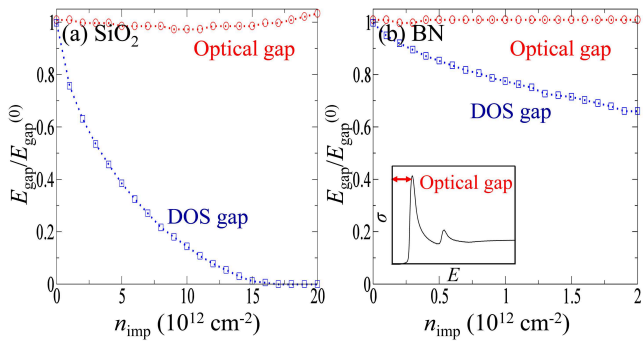}
	\caption{(Color online) DOS gap and optical gap in intrinsic BLG
	as a function of impurity density.
	Parameters in (a) correspond to an SiO$_2$ substrate and in (b) to a
	h-BN substrate \cite{substrates}. To aid comparison, the charged
	impurity concentration of an SiO$_2$ substrate is of the order of
	$10^{12}\,\mathrm{cm}^{-2}$ whereas for a h-BN substrate it is of
	the order of $10^{11}\,\mathrm{cm}^{-2}$.}
	\label{fig:gap}
\end{figure}

Following the experimental definition \cite{mak2009}, the optical gap is
defined as the energy at the top of the first peak in the optical
conductivity of intrinsic graphene.
This peak results from the direct optical excitations from the valence
to the conduction band and is the most direct way of
extracting the gap from an optical experiment. This gap is illustrated
in the inset to Fig. \ref{fig:gap}(b).

The effect of disorder on the DOS gap is easy to understand in Fig.
\ref{fig:dos}. With increasing impurity concentration, the gap size
reduces and eventually closes at a finite value of the impurity
concentration. This figure shows the DOS for a fairly
large potential asymmetry $u=0.2$ eV (which gives
$E_\mathrm{gap}^{(0)}=0.166$ eV), and a smaller gap generated by
a smaller asymmetry would be closed by a smaller concentration of
charged impurities.

\begin{figure}[tb]
	\includegraphics[width=0.98\linewidth]{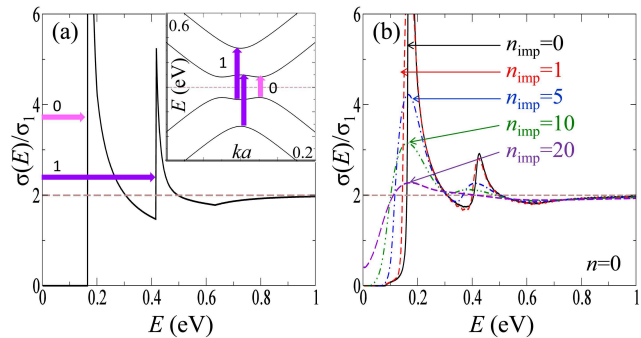}
	\caption{(Color online) (a) Interband optical conductivity of biased
	intrinsic BLG with $u=0.2$ eV in the absence of
	disorder.
	The inset shows the band structure with the interband transitions
	indicated by arrows.
	(b) Optical conductivity of disordered biased BLG for
	impurity densities $n_\mathrm{imp} =
	0,1,5,10,20\times10^{12}\,\mathrm{cm}^{-2}$.}
	\label{fig:optical-zeroEF}
\end{figure}

The effect of disorder on the optical gap is also easy to determine if
it is extracted from the optical conductivity of intrinsic BLG. 
Figure \ref{fig:optical-zeroEF}(a) shows the interband contribution to
the nondisordered optical conductivity of intrinsic BLG
\cite{min2009} along with a sketch showing
the origin of the two peaks. The transition labeled ``0'' is the one with
the lowest energy and corresponds to the excitation of carriers across the
band gap. Figure \ref{fig:optical-zeroEF}(b) shows the same quantity in
the presence of disorder. 
In addition to including the impurity scattering effects from
$n_\mathrm{imp}\neq 0$, we have included a finite phenomenological
level-broadening of $\eta = 1$ meV in the numerics 
which slightly rounds off the peaks in the $n_\mathrm{imp}=0$ case. 
The introduction of charged impurities further broadens the peaks
corresponding to the interband transitions but leaves their position
almost unchanged. This explains why the optical gap extracted from
experiment [which is defined by the peak value of $\sigma(\omega)$] is
very close to that predicted by tight-binding theory without disorder.

The DOS and optical gaps extracted from these calculations are
shown in Fig. \ref{fig:gap} as a function of the impurity
density \cite{substrates}.
The gaps are identical in the nondisordered system, but as the density
of charged impurities increases, the DOS gap is suppressed much more
quickly than the optical gap. In the majority of the original
experiments on transport in BLG, the samples are made with
exfoliated graphene on SiO$_2$ substrates and would therefore have had a
large amount of charged impurity disorder. 
We believe that this observation accounts for the difference in the
measured optical and transport gaps in this system.
The actual transport gap could be considerably smaller than our
theoretical DOS gap \cite{rossi2011}, as we have already noted.

\begin{figure}[tb]
	\includegraphics[width=0.98\linewidth]{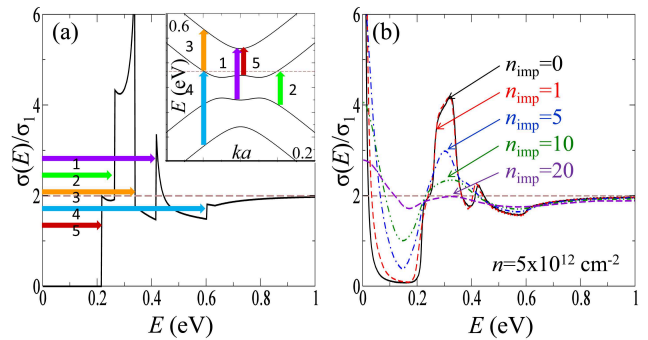}
	\caption{(Color online)
	(a) Interband optical conductivity of extrinsic biased BLG
	with carrier density $n=5\times 10^{12}\,\mathrm{cm}^{-2}$
	and $u=0.2$ eV in the absence of disorder.
	The inset shows the band structure with the interband transitions
	indicated by arrows and the dotted line shows the Fermi energy.
	(b) Optical conductivity of disordered biased BLG for
	several values of the impurity density.}
	\label{fig:optical-finEF}
\end{figure}

Figure \ref{fig:optical-finEF} shows the optical conductivity
calculated for extrinsic BLG on an SiO$_2$ substrate \cite{substrates}.
In this case, the extra transitions [which are sketched in the inset to 
Fig.~\ref{fig:optical-finEF}(a)] make
the optical conductivity a highly nontrivial function of energy and no
simple way of extracting an effective optical gap seems obvious.
The interband optical conductivity (calculated using the formula from
Ref. \onlinecite{min2009}) in the nondisordered case is shown in
Fig.~\ref{fig:optical-finEF}(a).
The inclusion of disorder [Fig~\ref{fig:optical-finEF}(b), using 
Eq.~\eqref{eq:conductivity}] further masks the structure and makes it
very difficult to reliably identify the peaks in the observed optical
spectra. 
The peak at low energy is due to the Drude contribution from intraband
processes near the Fermi surface.
For example, several experimental studies extract the size of the band
gap by comparing the energy of two peaks relating to different interband
transitions. In the language of Fig.  \ref{fig:optical-finEF}(a), these
transitions are labeled 1 (purple online) and 5 (red online). 
Making the identification with the plot in 
Fig.~\ref{fig:optical-finEF}(b), we see that
transition 1 corresponds to the second (smaller) peak, while transition
5 corresponds to the slight shoulder on the low energy side of the main
peak. As disorder increases, the peak due to transition 1 quickly
disappears, and the shoulder due to transition 5 becomes indistinct. It is
therefore impossible to use this method to extract the optical gap in
the presence of a realistic amount of charged impurity disorder in
BLG, and the peaks seen in the optical conductivity are not
obviously related to any individual transition. This is a qualitative
new finding of our work.

Finally, we discuss a practical way to test our theory and to bring the
DOS and optical gaps much closer in energy. To do this, optical
spectroscopy and charge transport experiments should be performed on the
same BLG device using an ultra-clean substrate, such as
hexagonal boron nitride (h-BN) which has an atomically smooth surface
and much lower impurity concentration than current SiO$_2$ devices
\cite{dean2010, dassarma2011}.
Recently, gated graphene layers on h-BN substrates have been
successfully fabricated \cite{dean2010}, and it was demonstrated that
graphene on h-BN substrates have substantially higher mobility than
graphene on SiO$_2$ substrates by an order of magnitude or more.

\begin{figure}[tb]
	\includegraphics[width=0.98\linewidth]{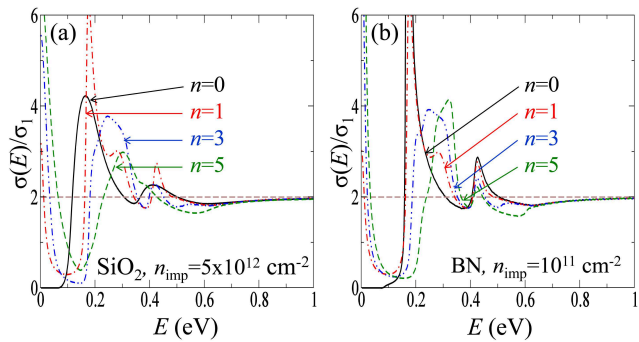}
	\caption{(Color online) Optical conductivity of biased BLG
	with $u=0.2$ eV on different substrates. (a)
	$n_{\rm imp}=5\times 10^{12}\,\mathrm{cm}^{-2}$ on SiO$_2$ 
	substrate and 
	(b) $n_{\rm imp}=10^{11}\,\mathrm{cm}^{-2}$ on an h-BN substrate. 
	In each case, the black line shows the intrinsic case, and the other
	lines show different carrier densities for extrinsic graphene,
	measured in $10^{12}\,\mathrm{cm}^{-2}$.}
	\label{fig:optical-EFvar}
\end{figure}

Figures \ref{fig:optical-EFvar}(a) and \ref{fig:optical-EFvar}(b) show 
the optical conductivity
for a typical impurity density of SiO$_2$ substrate and h-BN substrate
for a range of carrier densities. Apart from different disorder
strengths, the two substrates are distinguished by different dielectric
environments \cite{substrates}.
Because of the relatively low impurity density, the optical conductivity
for the h-BN substrate shows sharper peaks than those of SiO$_2$
substrate allowing for more accurate identification of the origin of the
various features and hence for a more accurate determination of the band
gap.
Figure \ref{fig:gap}(b) shows energy gaps extracted from the DOS and
optical conductivity of biased intrinsic BLG with
$u=0.2$ eV on an h-BN substrate as a function of impurity
density range. 
In contrast to the same plots for SiO$_2$ shown in Fig. \ref{fig:gap}(a), 
the deviation of the DOS gap from the optical gap for the h-BN substrate
is rather small. Note that experimentally realistic values of the
impurity density on the h-BN substrate \cite{dassarma2011} are of the
order of $10^{11}\,\mathrm{cm}^{-2}$ so that the DOS gap is
approximately 95\% of the optical gap. Therefore, BLG
mounted on an h-BN
substrate can be used as a tunable gap semiconductor and will yield a
significantly higher on/off ratio than the corresponding SiO$_2$
substrate.

In summary, using the SCBA, we have shown that optical experiments and
transport measurements do not always probe the same gap in biased BLG.
The disorder affects the DOS so that the gap is reduced,
while the position of the peak associated with interband transitions in
the optical conductivity of intrinsic BLG is rather robust against
disorder. In contrast, the features in the optical conductivity of
extrinsic BLG which are compared to extract the band gap become
indistinct with significant disorder.  Our work clearly establishes the
danger of extracting a na\"ive optical band gap in bilayer graphene
from the optical absorption data without a thorough theoretical
analysis, calling into question several recent experimental claims
\cite{zhang2009, li2009, mak2009}.
We also suggest that samples with low disorder (such as BLG on an h-BN
substrate, or suspended BLG) should show comparable DOS and optical
gaps.

The work is supported by US-ONR and NRI-SWAN.



\begin{thebibliography}{999}
\bibitem{substrates}
We assume that the bilayer graphene samples have a single gate so that
the dielectric environment of the graphene is the average of the gate
and air. With $\epsilon_{\mathrm{SiO_2}}=4$ and
$\epsilon_{\mathrm{h-BN}} = 7$, the effective dielectric constants are
$2.5$ and $4$ respectively.
\end{thebibliography}
\end{document}